\documentclass[11pt]{article}
\usepackage{moriond,epsfig}

\def\Journal#1#2#3#4{{#1} {\bf #2}, #3 (#4)}

\def\NPB{{\em Nucl. Phys.} B}
\def\PLB{{\em Phys. Lett.}  B}
\def\PRL{\em Phys. Rev. Lett.}
\def\PRD{{\em Phys. Rev.} D}
\def\PTP{\em Prog. Theor. Phys}
\def\ZPC{{\em Z. Phys.} C}
\def\EPJ{{\em Eur.Phys.J} C}

\def\babar{\mbox{\sl B\hspace{-0.4em} {\small\sl A}\hspace{-0.37em} \sl B\hspace{-0.4em} {\small\sl A\hspace{-0.02em}R}}}
\def\ckmfitter{{\em CKMfitter}}
\def\rar{\rightarrow}

\def\be{\begin{equation}}
\def\ee{\end{equation}}
\def\bea{\begin{eqnarray}}
\def\eea{\end{eqnarray}}

\def\rfit{{\em R}fit}

\def\eg{{\it e.g.}}
\def\ea{{\it et al.}}

\def\rs{\raisebox{1.3ex}[-1.5ex]}
\def\epsk{\epsilon_K}
\def\dmd{\Delta m_d}
\def\dms{\Delta m_s}
\def\etabar{\bar{\eta}}
\def\rhobar{\bar{\rho}}
\def\stbwa{{\rm sin}2\beta_{\rm WA}}
\def\fbd{f_{B_d}}
\def\MSbar{\overline{\rm MS}}
\def\sta{{\rm sin}2\alpha}
\def\stb{{\rm sin}2\beta}

\def\pp{$B \rightarrow \pi\pi$~}
\def\kp{$B \rightarrow K\pi$~}
\def\ifb{$fb^{-1}$~}
\def\Cpipi{C_{\pi\pi}}
\def\Spipi{S_{\pi\pi}}
\def\bdbar{B_d^0 {\overline B}_d^0}

\newcommand{\MSbm}{{\overline{\rm MS}}}
\begin{document}
\vspace*{4cm}
\title{THE CKM PARADIGM: IMPLICATIONS OF THE MOST RECENT RESULTS}
\author{SANDRINE~LAPLACE}
\address{Laboratoire de l'Acc\'el\'erateur Lin\'eaire, IN2P3-CNRS et Universit\'e Paris-Sud,
BP 34, F-91898 Orsay Cedex, FRANCE }

\maketitle\abstracts{The implications of the most recent experimental
results in the $B$ and $K$ meson systems on the CKM paradigm are investigated
by means of a global fit to the theoretical predictions of the 
Standard Model. We advocate the frequentist approach \rfit, which is implemented in the
\ckmfitter~package. Within this approach, constraints on the 
CKM parameters and other quantities of interest are obtained, and extensions of
the Standard Model are investigated.
\begin{flushright} LAL 02-79 \end{flushright} }
\section{Introduction}
The electroweak sector of the Standard Model (SM) has been intensively studied during
the last decades, in particular at LEP and SLD. $CP$ violation in flavour changing processes 
is the next sector to be investigated. $CP$ violation was first observed in the 
Kaon system~\cite{Kaons}, and recently established in the $B$-meson system~\cite{BCPV}, 
but within the SM it cannot produce a baryon asymmetry as large as what is observed in the
universe~\cite{matterasym}. A variety of models beyond the SM introduce new sources of 
$CP$ violation: therefore, $CP$ violation is an interesting place to look for new physics, 
and particularly in the $B$-meson system where there are numerous large $CP$-violating observables.

Within the Standard Model, $CP$ violation is accounted for by one phase~\footnote{With more than three
generations, there would be more CP violating phases.} in the CKM matrix~\cite{Cabibbo,KM} 
which describes the charged current couplings among left-handed quark fields.
With three generations, the CKM matrix is parameterized by four parameters (including
the $CP$ violating phase). We use the Wolfenstein parameterization~\cite{Wolfenstein},
which, following the observed hierarchy between the matrix elements, expands the 
CKM matrix in terms of the four parameters $\rho$, $\eta$, $A$ and $\lambda$ ($\lambda \simeq 0.22$ being
the expansion parameter). Presently, $\lambda$ is known at the $1 \%$ level, 
$A$ at the $5 \%$ level, whereas $\rho$ and $\eta$ are the least known CKM parameters.
Many relations between the matrix elements follow from the unitarity of the CKM matrix.
In particular, the unitarity relation linking the first and third rows of the matrix
is usually displayed as the Unitarity Triangle (UT) in the complex plane 
($\rhobar$,$\etabar$)~\footnote{The replacements $\rho \rightarrow \rhobar=\rho(1-\lambda^2/2)$ and
$\eta \rightarrow \etabar=\eta(1-\lambda^2/2)$ improves the accuracy of the UT apex.}.

Investigating the charged current couplings among quarks is not a straightforward task since
quarks also (and mainly) undergo the strong interaction, in which calculations are not
always accurate. Therefore, the SM prediction 
of many CKM observables suffers from uncertainties arising from the use of approximations 
in QCD computations. These hard-to-quantify uncertainties do not belong to the well-defined 
framework of Gaussian statistics. 

In order to avoid drawing wrong conclusions from the global CKM fit of such observables, 
one needs a robust and sufficiently conservative approach: we advocate the frequentist 
approach \rfit~\cite{EPJCPaper}, within the framework of the \ckmfitter~package, and describe its 
main features and results in the following.
\section{Statistical Framework and Fit Ingredients}
In the \rfit~frequentist statistical framework, theoretical uncertainties are treated
without {\it a priori} knowledge except for the definition of ranges~\cite{BabarBook,MHS}.
In particular, no probability distribution functions are considered~\cite{Bayesians}.
Practically, the fit likelihood is the product
of an experimental part, ${\cal L}_{\rm exp}(x_{\rm exp}-x_{\rm theo}(y_{\rm mod}))$
comparing the measurements $x_{\rm exp}$ to the theoretical predictions $x_{\rm theo}$
(the latter depending on some model parameters $y_{\rm mod})$~\footnote{The four CKM parameters  
are part of these model parameters.}, with a theoretical part,
${\cal L}_{\rm theo}(y_{\rm QCD})$~\footnote{A range of $y_{\rm QCD}$ is obtained from QCD theoretical
calculations: ${\cal L}_{\rm theo}(y_{\rm QCD})$=1 for $y_{\rm QCD}$
inside the range, and 0 when outside.} which describes our knowledge on the QCD parameters
$y_{\rm QCD} \in y_{\rm mod}$ . A pseudo-$\chi^2$ is built as $\chi^2=-2ln{\cal L}$ and
is minimized in the fit.

The analysis is divided into three steps: first, the overall consistency between data and the
theoretical framework (here, the SM) is tested by means of a Monte Carlo simulation.
If the agreement between data and the SM is found to be acceptable, confidence levels (CL)
in parameter subspaces are determined (\eg, in the $\rhobar$-$\etabar$ plane). Finally,
extensions of the SM can be tested and limits on new physics parameters are determined.

Table~\ref{tab:smtab} gives the input ingredients of the fit. For $|V_{ub}|$, the inclusive measurement
of LEP~\cite{VubLEPincl} and exclusive measurement of CLEO~\cite{VubCLEOexcl} have been used, as well
as the inclusive CLEO analysis using the $B \rightarrow X_s \gamma$ spectrum~\cite{VubCLEOincl}. 
The input to the fit is the product of these three measurement likelihoods. The inclusive and exclusive 
LEP and CLEO $|V_{cb}|$ measurements are averaged before entering the fit~\cite{HeikoHF9}.
For $\dms$, we translate the preferred value of the amplitude spectrum~\cite{lepbosc} into a confidence
level curve~\cite{CKMWorkshop}. Finally, we use the world average of $sin 2\beta$, dominated by the measurements of 
\babar\ and Belle~\cite{BCPV}.
\begin{table}[h]
\caption{
        Input observables and parameters for the global CKM fit. 
	When two errors are quoted, the first one 
	is statistical and the second one systematic. When one error is
	quoted, it is statistical unless stated otherwise.}\label{tab:smtab}
\vspace{0.4cm}
\begin{center}
    {\small
      \begin{tabular}{lc}\hline \\
        \rs{Parameter}        & \rs{Value $\pm$ Error(s)} \\
        \hline & \\
        \rs{$|V_{ud}|$}       &
        \rs{$0.97394\pm0.00089$} \\
        \rs{$|V_{us}|$}       &
        \rs{$0.2200\pm0.0025$} \\
        \rs{$|V_{ub}|$ (LEP incl.)~\cite{VubLEPincl}}  &
        \rs{$(4.08 \pm 0.61 \pm 0.47) \times 10^{-3}$} \\
        \rs{$|V_{ub}|$ (CLEO incl.)~\cite{VubCLEOincl}} &
        \rs{$(4.08 \pm 0.56 \pm 0.40) \times 10^{-3}$} \\
        \rs{$|V_{ub}|$ (CLEO excl.)~\cite{VubCLEOexcl}} &
        \rs{$(3.25 \pm 0.29 \pm 0.55) \times 10^{-3}$} \\
        \rs{$|V_{cd}|$}       &
        \rs{$0.224\pm0.014$} \\
        \rs{$|V_{cs}|$}       &
        \rs{$0.969 \pm 0.058$} \\
        \rs{$|V_{cb}|$~\cite{HeikoHF9}}   &
        \rs{$(40.4\pm 1.3 \pm 0.9)\times10^{-3}$} \\
        \hline
        & \\
        \rs{$|\epsk|$}        &
        \rs{$(2.271\pm0.017)\times10^{-3}$} \\
        \rs{$\dmd$~\cite{lepbosc}}  &
        \rs{$(0.496 \pm 0.007)~{\rm ps}^{-1}$} \\
        \rs{$\dms$~\cite{lepbosc}}  &
        \rs{See Text} \\
        \rs{$\stbwa$~\cite{BCPV}} & \rs{$0.780 \pm 0.077$} \\
        \hline
        & \\
        \rs{$m_c$}              & \rs{$(1.3\pm0.1_{syst})~{\rm GeV}$} \\
        \rs{$m_t(\MSbar)$}      & \rs{$(166.0\pm5.0)~{\rm GeV}$} \\
        \rs{$m_K$}              & \rs{$(493.677\pm0.016)~{\rm MeV}$} \\
        \rs{$\Delta m_K$}       & \rs{$(3.4885 \pm 0.0008)\times 10^{-15}~{\rm GeV}$} \\
        \rs{$m_{B_d}$}  & \rs{$(5.2794\pm0.0005)~{\rm GeV}$} \\
        \rs{$m_{B_s}$}  & \rs{$(5.3696\pm0.0024)~{\rm GeV}$} \\
        \rs{$m_W$}              & \rs{$(80.419\pm0.056)~{\rm GeV}$} \\
        \rs{$G_F$}              & \rs{$(1.16639\pm0.00001)\times 10^{-5}~{\rm GeV^{-2}}$} \\
        \rs{$f_K$}              & \rs{$(159.8\pm1.5)~{\rm MeV}$} \\
        \hline 
        & \\
        \rs{$B_K$}            &
        \rs{$0.87\pm0.06\pm0.13$} \\
        \rs{$\eta_{cc}$}      &
        \rs{$1.38\pm0.53_{syst}$} \\
        \rs{$\eta_{ct}$}      &
        \rs{$0.47\pm0.04_{syst}$} \\
        \rs{$\eta_{tt}$}      &
        \rs{$0.574\pm0.004_{syst}$} \\
        \rs{$\eta_B(\MSbm)$} &
        \rs{$0.55\pm0.01_{syst}$ } \\
        \rs{$\fbd\sqrt{B_d}$} &
        \rs{$(230\pm28\pm28)~{\rm MeV}$} \\
        \rs{$\xi$}            &
        \rs{$1.16 \pm 0.03 \pm 0.05$} \\
        \hline
      \end{tabular}
      }
\end{center}
\end{table}
\section{Standard Fit}
The consistency between data and the SM is gauged by the value of the overall minimum $\chi^2$, 
from which a CL is inferred by means of a Monte Carlo simulation. Many sets of measurements are generated 
following the experimental likelihood distribution in which the theoretical predictions are computed
for the best fitted values of the theoretical parameters.
A fit is then performed on each of these sets, varying all the model parameters freely, and the
CL of the SM fit is assigned by integrating the $\chi^2$ distribution thus obtained, up to the overall minimum $\chi^2$.

\begin{figure}[h]
\begin{center}
\psfig{figure=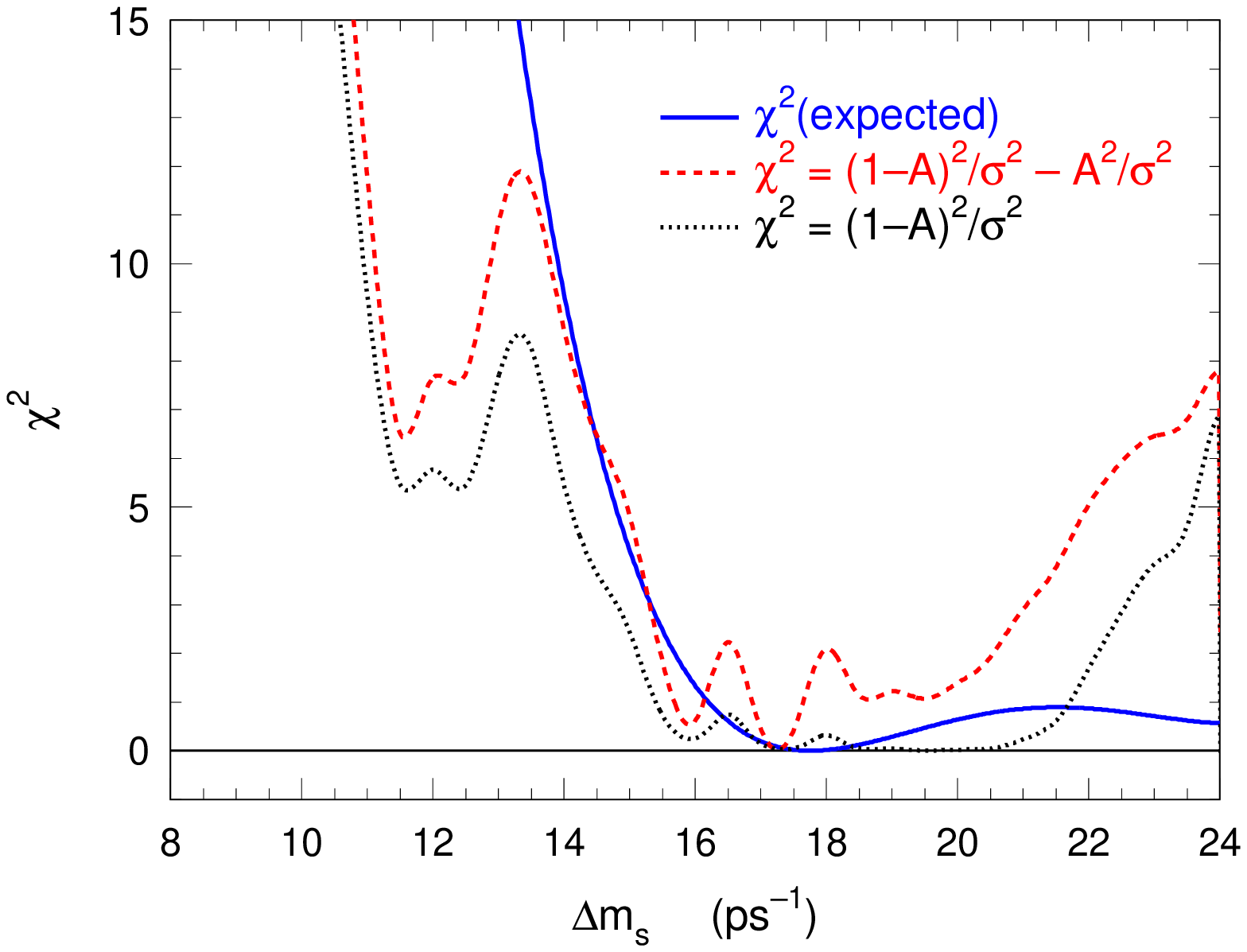,height=2.2in}
\psfig{figure=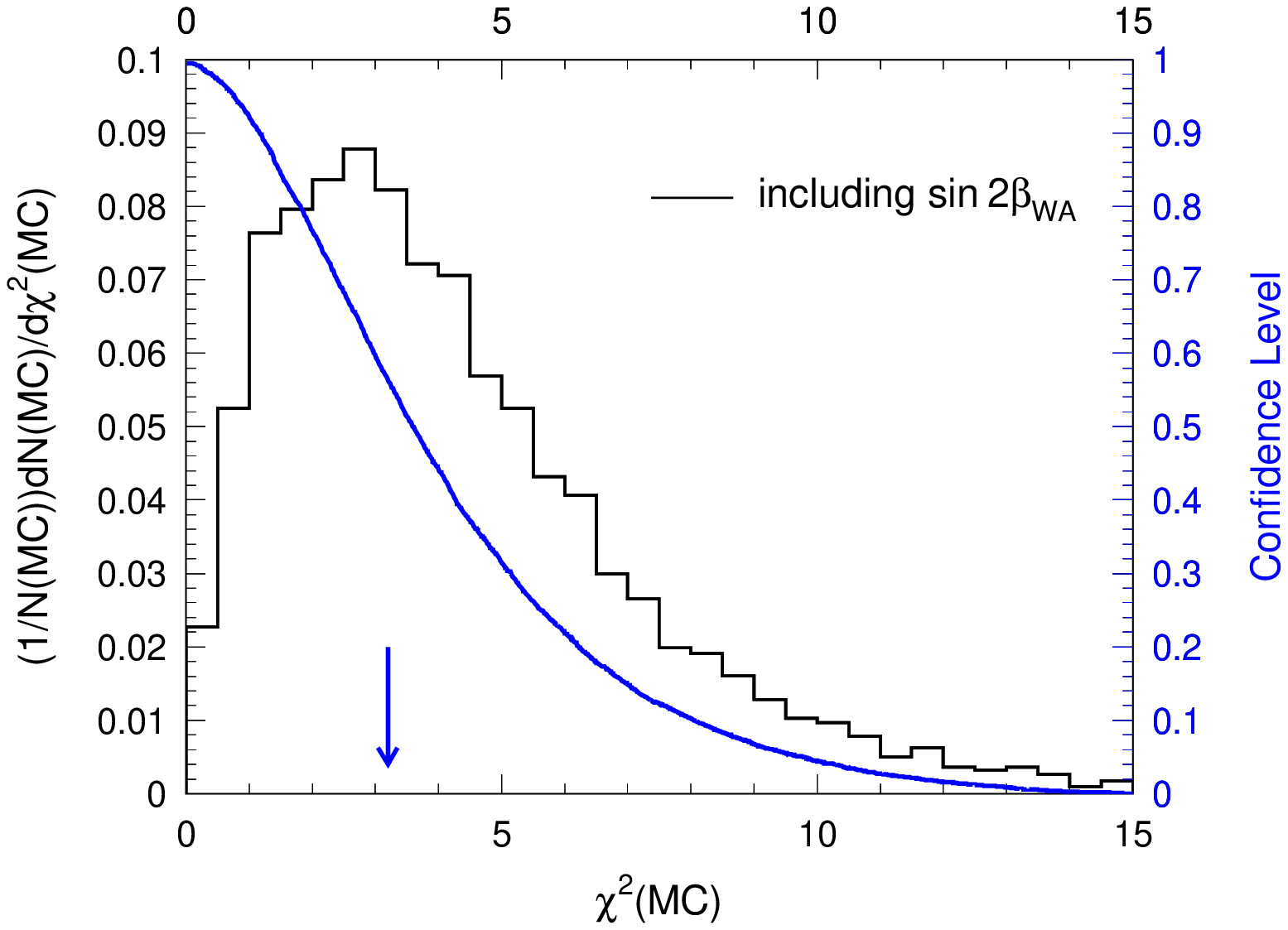,height=2.2in}
\end{center}
\caption{\underline{Left figure:} $\chi^2$ curves for various interpretations of the
amplitude spectrum of $\dms$: the modified amplitude method~\protect\cite{EPJCPaper} (dotted line),
the likelihood ratio method~\protect\cite{Bayesians} (dash-dotted line), the 
expected $\chi^2$ curve for the current preferred value of $\dms$~\protect\cite{CKMWorkshop} (solid line). 
\underline{Right figure:} 
Test of the goodness of the SM fit. The histogram (corresponding to the left scale)
is the Monte Carlo-generated $\chi^2$ distribution (see text), which is integrated to compute
a CL, as given by the smooth curve (right scale). The $\chi^2$ value of the current SM fit is indicated
by the arrow, corresponding to a CL of $57 \%$.
\label{fig:chi2min}}
\end{figure}

\begin{figure}[h]
\begin{center}
\psfig{figure=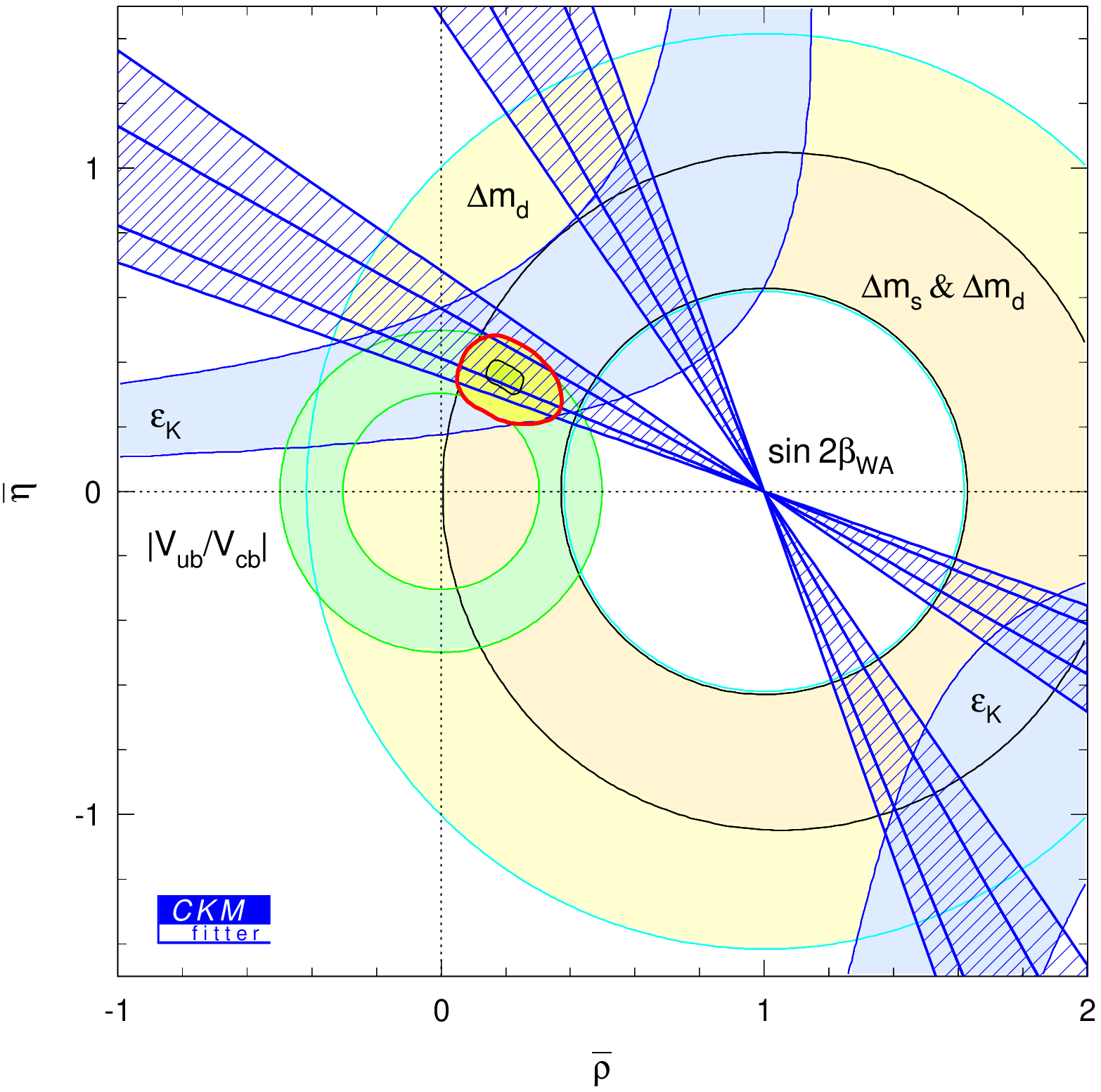,height=3.in}
\psfig{figure=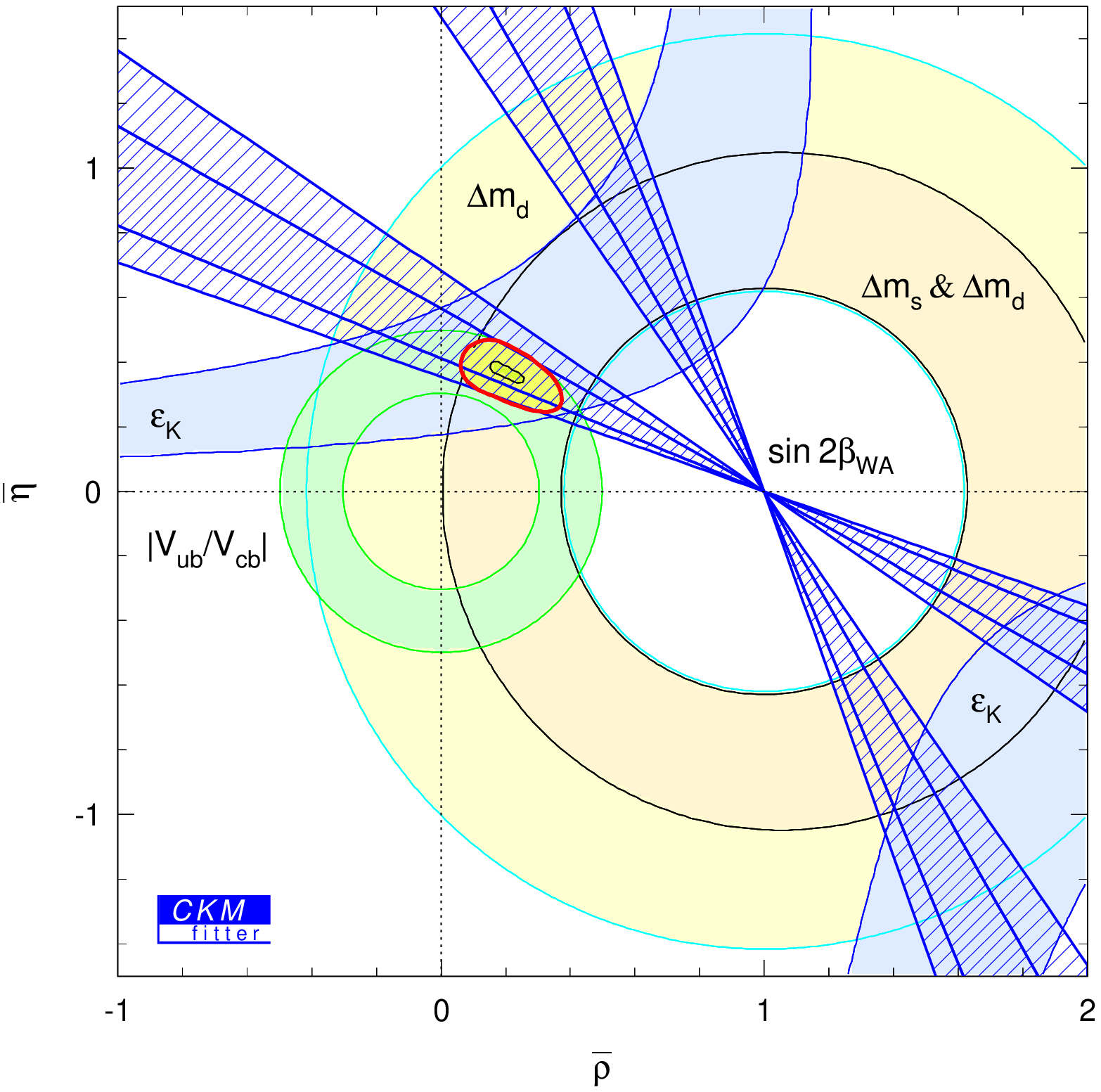,height=3.in}
\caption{Constraints in the $\rhobar-\etabar$ plane: regions with more than
$5 \%$ CL are shown for the main observables: $\dmd,~\dms,~\epsk,~|V_{ub}/V_{cb}|$.
The global SM fit constraint lies at the intersection of all individual measurements. 
For both the global fit and the new world average $sin 2\beta$ constraint, two contour lines
including points with more than $5 \%$ and $32 \%$ CL are drawn. 
\underline{Left plot:} the $sin 2\beta$ constraint is not included in the global SM fit, but
is simply overlaid: the measurement is in perfect agreement with the SM expectation.
\underline{Right plot:} the $sin 2\beta$ constraint is included in the global SM fit, 
and allows to pin down our knowledge on $\rhobar-\etabar$.
\label{fig:rhoeta}}
\end{center}
\end{figure}

Figure~\ref{fig:chi2min} shows the distribution of the Monte Carlo generated $\chi^2$ (histogram),
and the associated CL (smooth curve). The arrow shows the $\chi^2$ of the standard fit, corresponding to 
a CL of $57 \%$. 
The agreement between data and Standard Model is thus excellent, and we assign CL to
various parameters of interest: Fig.~\ref{fig:rhoeta} displays CL in the
($\rhobar$-$\etabar$) plane. For the individual constraints are shown the regions inside 
which $CL \ge 5\%$, including the $4$-fold ambiguity from the $\sin 2 \beta$ measurement. 
The latter is in perfect agreement with the SM prediction, demonstrating that the CKM mechanism
is the dominant source of $CP$ violation in flavour changing processes at the electroweak scale.
%
\begin{table}[t]
\begin{center}
{\small
\setlength{\tabcolsep}{0.83pc}
\begin{tabular}{lc} \hline
& \\[-0.3cm]
Parameter	& $95\%$ CL region		\\[0.2cm]
\hline
& \\[-0.3cm]
$\lambda$	& $0.2221\pm0.0041$		\\
$A$		& 0.76 - 0.90			\\
$\rhobar$	& 0.08 - 0.35 			\\
$\etabar$	& 0.28 - 0.45			\\
$J$ $[10^{-5}]$	& 2.2 - 3.5			\\
$\sta$		& $-0.81$ - 0.43 		\\
$\stb$		& 0.64 - 0.84			\\
$\alpha$	& $77^\circ$ - $117^\circ$	\\
$\beta$		& $19.9^\circ$ - $28.6^\circ$	\\
$\gamma=\delta_{CP}$	
		& $40^\circ$ - $78^\circ$	\\[0.2cm]
\hline
\end{tabular}
\hspace{0.5cm}
\begin{tabular}{lc} \hline
& \\[-0.3cm]
Parameter	& $95\%$ CL region		\\[0.2cm]
\hline
& \\[-0.3cm]
$|V_{ud}|$	& $0.97504\pm0.00094$	 \\
$|V_{ub}|$ $[10^{-3}]$
		& 3.15 - 4.37	\\
$|V_{cb}|$ $[10^{-3}]$
		& 36.9 - 43.6	\\
$|V_{td}|$ $[10^{-3}]$
		& 6.3 - 9.1	\\
$|V_{ts}|$ $[10^{-3}]$
		& 36.4 - 43.0	\\
$|V_{tb}|$	& 0.99905 - 0.99932	\\
${\rm BR}(K^0_{\rm L}\rightarrow\pi^0\nu\bar\nu)$ $[10^{-11}]$
		& 1.6 - 4.2	\\
${\rm BR}(K^+\rightarrow\pi^+\nu\bar\nu)$ $[10^{-11}]$
		& 5.1 - 8.4 	\\
${\rm BR}(B^+\rightarrow\tau^+\nu_\tau)$ $[10^{-5}]$
		& 7.2 - 22.1	\\
${\rm BR}(B^+\rightarrow\mu^+\nu_\mu)$ $[10^{-7}]$
		& 2.9 - 8.7	\\[0.2cm]
\hline
\end{tabular}}
\caption[.]{
	Fit results for the CKM parameters, the CKM
	matrix elements and branching ratios of some rare $K$ and $B$
	meson decays, {\em including the world average $\stb$ in the fit}.
	Ranges are given for the quantities that are limited by 
	systematic theoretical errors.}
\label{tab:results}
\end{center}
\end{table}
\section{Beyond the Standard Fit}
\subsection{Rare $K$ Decays}
The detection of a second $K^+ \rightarrow \pi^+\nu\bar{\nu}$ event was announced
early 2002 by the E787 Collaboration at Brookhaven\cite{E787}, leading to a branching ratio
of $(1.57^{+1.75}_{-0.82})\times10^{-10}$ which is compatible with the
SM prediction~\cite{RareK} given in Table~\ref{tab:results}.
A total of 5 to 10 SM events is expected to be detected at the successor experiment E949 at BNL~\cite{E949}, 
and thus this channel will become an important ingredient of the CKM fits.
\subsection{Charmless $B$ Decays}
Both the $\alpha$ and $\gamma$ angles of the UT can in principle be extracted from the charmless two-body decays 
\pp and \kp, for which the data start to accumulate. Therefore, a lot of effort~\cite{2body,BBNS}
has been put recently in the calculation of the charmless two-body hadronic
matrix elements using the factorization concept.

One of these approaches, the QCD factorization approach (QCD FA)~\cite{BBNS}, has been implemented in \ckmfitter. The results
obtained from these calculations are labelled by ``R$\&$D'', meaning that they
are still subject to theoretical questions and that we do not intend
to infer constraints on the CKM parameters from them. Rather, we intend to test
the consistency of the QCD FA predictions in a global fit.

In the QCD FA, the tree, strong penguin and electroweak penguin amplitudes are computed, 
the annihilation diagrams estimated, and the soft physics contribution in the spectator interaction are
parameterized. The strong phases are found to be small, and thus QCD FA predicts small direct
$CP$ asymmetries.
\subsubsection{Global Fit of Branching Ratios and $CP$ Asymmetries}
In a first stage, a global fit on QCD FA is performed using the recent charmless two-body
branching ratios (BR), given in table~\ref{tab:BF}, and time-integrated 
$CP$ asymmetries ($A_{\rm CP}$), given in Table~\ref{tab:acp}, as measured by the \babar\ , Belle and 
CLEO collaborations~\cite{BR_ACP}. 
\begin{table}[t]
\caption{Branching ratios of the charmless two-body modes used in the global fit. The
measurements from CLEO, \babar\ and Belle collaborations~\protect\cite{BR_ACP} have been averaged
into so-called World Average (WA) values.
\label{tab:BF}}
\vspace{0.4cm}
\begin{center}
\begin{tabular}{|c|c|c|c|c|}
\hline
BR ($\times 10^{-6}$)  & CLEO (9.1 \ifb) & \babar\ (55.6 \ifb) & Belle (31.7 \ifb) & WA \\ \hline
$B^0 \rightarrow \pi^+\pi^-$ & $4.3^{+1.6}_{-1.4} \pm 0.5$ & $5.4\pm 0.7\pm 0.4$      & $5.1\pm 1.1\pm 0.4$    & $5.17\pm 0.62$ \\
$B^0 \rightarrow K^+\pi^-$   & $17.2^{+2.5}_{-1.4}\pm 1.2$ & $17.8\pm 1.1\pm 0.5$     & $21.8\pm 1.8\pm 1.5$   & $18.5 \pm 1.0$ \\
$B^0 \rightarrow K^+K^-$     & $< 1.9 (90 \%)$            & $< 2.5 (90 \%)$         & $< 2.7 (90 \%)$       &  \\
$B^+ \rightarrow \pi^+\pi^0$ & $5.6^{+2,6}_{-2.3}\pm 1.7$  & $5.1\pm 2.0\pm 0.8$      & $7.0\pm 2.2\pm 0.8$    & $5.9 \pm 1.4$ \\
$B^+ \rightarrow K^+\pi^0$   & $11.6^{+3.0}_{-2.7}{}^{+1.4}_{-1.3}$ & $10.8\pm 2.1 \pm 1.0$ & $12.5\pm 2.4\pm 1.2$ & $11.6^{1.6}_{-1.5}$ \\
$B^+ \rightarrow K^0\pi^+$   & $18.2^{4.6}_{-4.0}\pm 1.6$  & $18.2\pm 3.3\pm 2.2$     & $18.8\pm 3.0\pm 1.5$   & $18.5^{2.3}_{-2.2}$ \\
$B^0 \rightarrow K^0\pi^0$ & $14.6^{5.9}_{5.1}{}^{2.4}_{-3.3}$ & $8.2 \pm 3.1\pm 1.2$ & $7.7\pm 3.2\pm 1.6$    & $9.0\pm 2.2$ \\
\hline
\end{tabular}
\end{center}
\end{table}

\begin{figure}[h]
\begin{center}
\psfig{figure=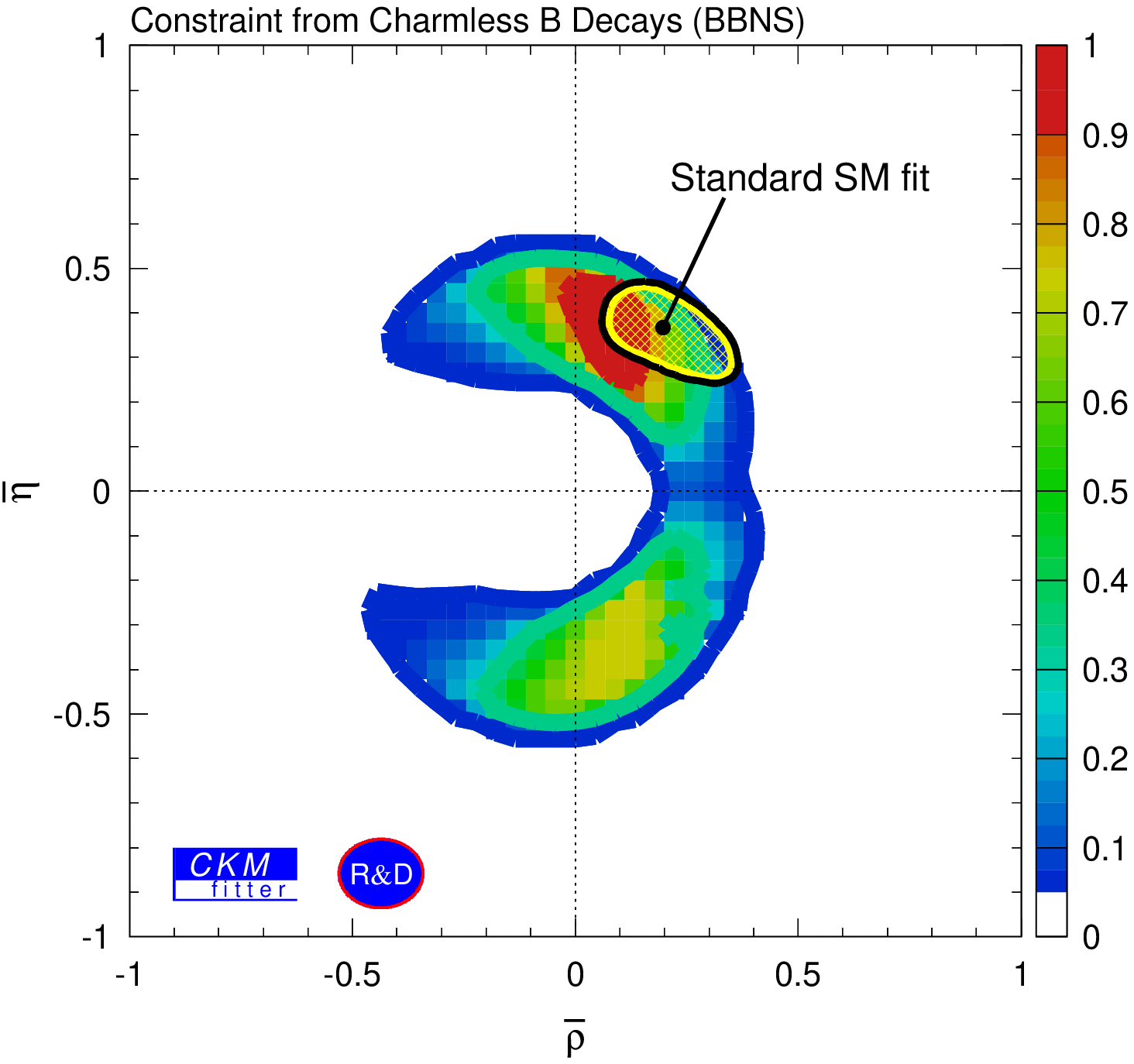,height=3.18in}
\psfig{figure=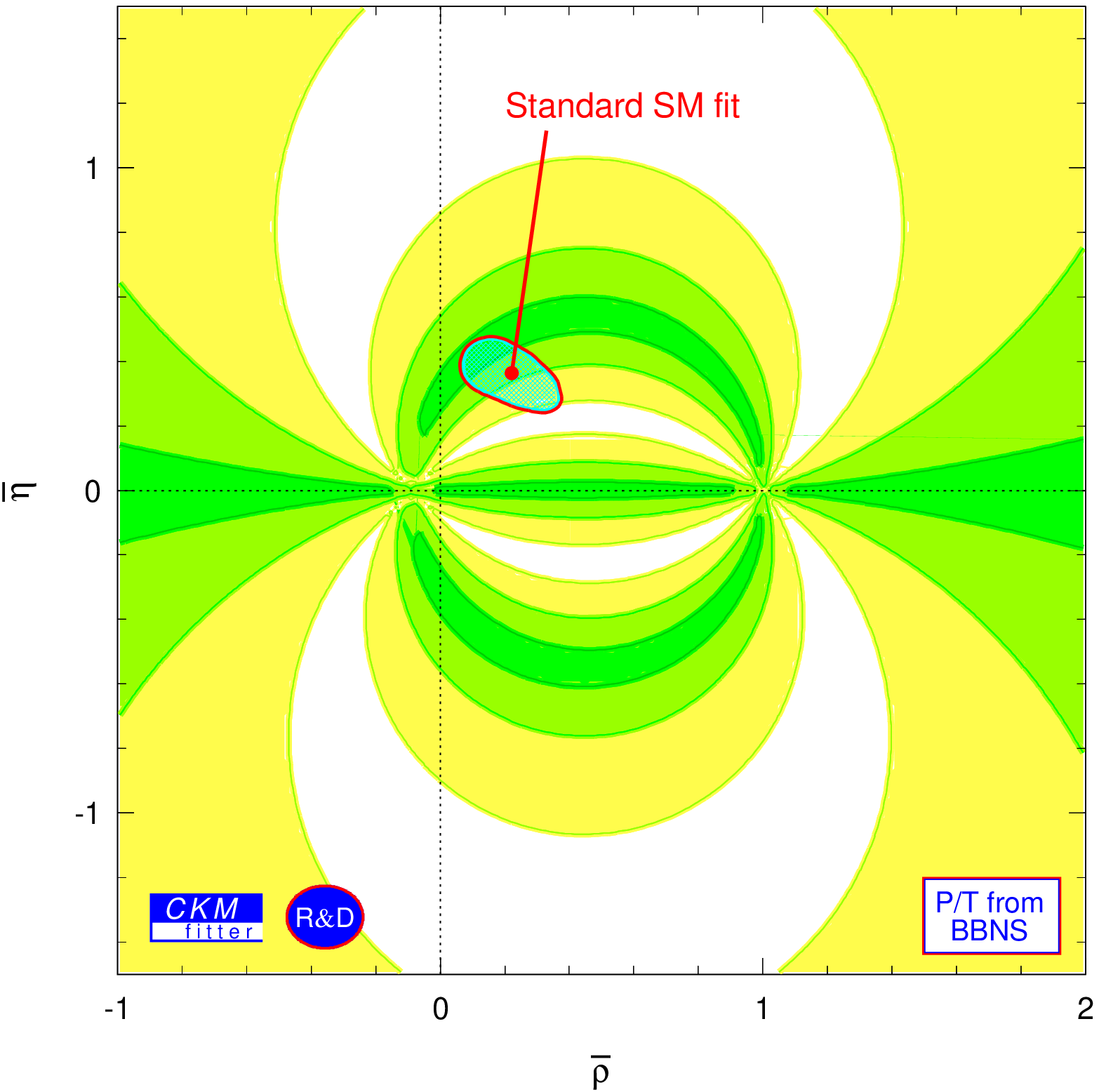,height=2.9in}
\caption{\underline{Left Plot}: CL in the ($\rhobar-\etabar$) plane obtained from a fit using
branching ratios and $CP$ asymmetries from Table~\ref{tab:BF}, in the framework of the QCD FA approach.
\underline{Right Plot}: CL in the ($\rhobar-\etabar$) plane obtained from a fit using $\Cpipi$ and
$\Spipi$ from \babar\, where $\delta$ and the ratio $|P/T|$ are predicted by the QCD FA approach.
The light, medium and dark shade zones include points with more than $5 \%$,$32 \%$,$90 \%$ CL. \label{fig:acp}}
Also shown is the $95 \%$ CL zone of the SM fit.
\end{center}
\end{figure}
\begin{table}[t]
\caption{Time-integrated direct $CP$ asymmetries $A$ in $B\rar K\pi$ and $CP$ parameters $\Cpipi$
and $\Spipi$~\protect\cite{BR_ACP} of the time-dependent analysis in $B^0 \rar \pi^+\pi^-$.
\label{tab:acp}}
\vspace{0.4cm}
\begin{center}
\begin{tabular}{|c|c|}
\hline
Observable  & Value \\ \hline
$A_{\rm CP}(B^0 \rightarrow K^+\pi^-)$ & $-0.05\pm 0.05$ (WA)\\
$A_{\rm CP}(B^+ \rightarrow K^+\pi^0)$ & $-0.09\pm 0.12$ (WA)\\
$A_{\rm CP}(B^+ \rightarrow K^0\pi^+)$ & $0.18 \pm 0.10$ (WA)\\
\hline
$\Spipi$		      & $-0.01\pm0.38$ (\babar) \\
		 	      & $-1.21^{\,+0.41}_{\,-0.30}$ (Belle)\\
$\Cpipi$	 	      & $-0.02\pm0.30$ (\babar)\\
       			      & $-0.94^{\,+0.32}_{\,-0.27}$ (Belle)\\
\hline
\end{tabular}
\end{center}
\end{table}
The numerous theoretical parameters are free to vary in the global fit according to 
the predications of Ref.~\cite{BBNS}. In the left plot of Fig.~\ref{fig:acp}, the QCD FA constraints 
from the charmless two-body BR and $A_{\rm CP}$ are compared
to the SM constraints coming from all the measurements of Table~\ref{tab:smtab}. 

One concludes that the QCD FA can accommodate the branching ratios and CP asymmetries 
of the \pp and \kp decays, and that the preferred value of the angle $\gamma$ is around $80^\circ$.
The constraints in the $\rhobar$-$\etabar$ plane are in agreement with those obtained in the standard fit.
\subsubsection{$\Cpipi$ and $\Spipi$ Fits using QCD FA}
The time-dependent CP asymmetry of $B^0 \rightarrow \pi^+\pi^-$ rate reads
\begin{equation}
   a_{CP}(t) = \Spipi {\rm sin}(\dmd \Delta t) - \Cpipi {\rm cos}(\dmd \Delta t)~,
\end{equation}
with $\Delta t$, the time difference between the $CP$  and the tag $B$ decays, and with the coefficients
given by
\begin{equation}
\label{eq:lambdaCP}
   \Spipi = \frac{2{\rm Im}\lambda_{\pi\pi}}{1 + |\lambda_{\pi\pi}|^2}
	\hspace{0.5cm} {\rm and} \hspace{0.5cm}
   \Cpipi = \frac{1 - |\lambda_{\pi\pi}|^2}{1 + |\lambda_{\pi\pi}|^2}~.
\end{equation}
The $CP$ parameter $\lambda_{\pi\pi}$ is given by 
\begin{equation}
\label{eq:lambda}
	\lambda_{\pi\pi} 
	= e^{-2i\beta} \frac{A(\overline B^0\rar\pi^+\pi^-)}
                            {A(B^0\rar\pi^+\pi^-)}
	=
	e^{2i\alpha}\frac{1 - (R_t/R_u) |P/T| e^{-i(\alpha-\delta)}}
			  {1 - (R_t/R_u) |P/T| e^{+i(\alpha+\delta)}}
	\equiv
	|\lambda_{\pi\pi}|e^{2i\alpha_{\rm eff}}~,
\end{equation}
where the phase $e^{-2i\beta}$ arises from $\bdbar$ mixing,
$R_{t(u)}$ are the sides of the UT~\cite{EPJCPaper}, and $T$, $P$ are the tree and
penguin amplitudes.
In the presence of penguin contributions, the UT angle
$\alpha$ enters the expression of $\lambda_{\pi\pi}$ with a correction
term which depends, in particular, on the relative strong phase 
$\delta \equiv {\rm arg}(P {T}^*)$ between the 
penguin and the tree amplitudes. As a consequence, a measurement
of the parameters $\Spipi$ and $\Cpipi$ cannot be translated
into $\sta$ without information on the relative strong phase
$\delta$ and the ratio $|P/T|$, \eg, as computed
by the QCD FA approach.

\begin{figure}[h]
\begin{center}
\psfig{figure=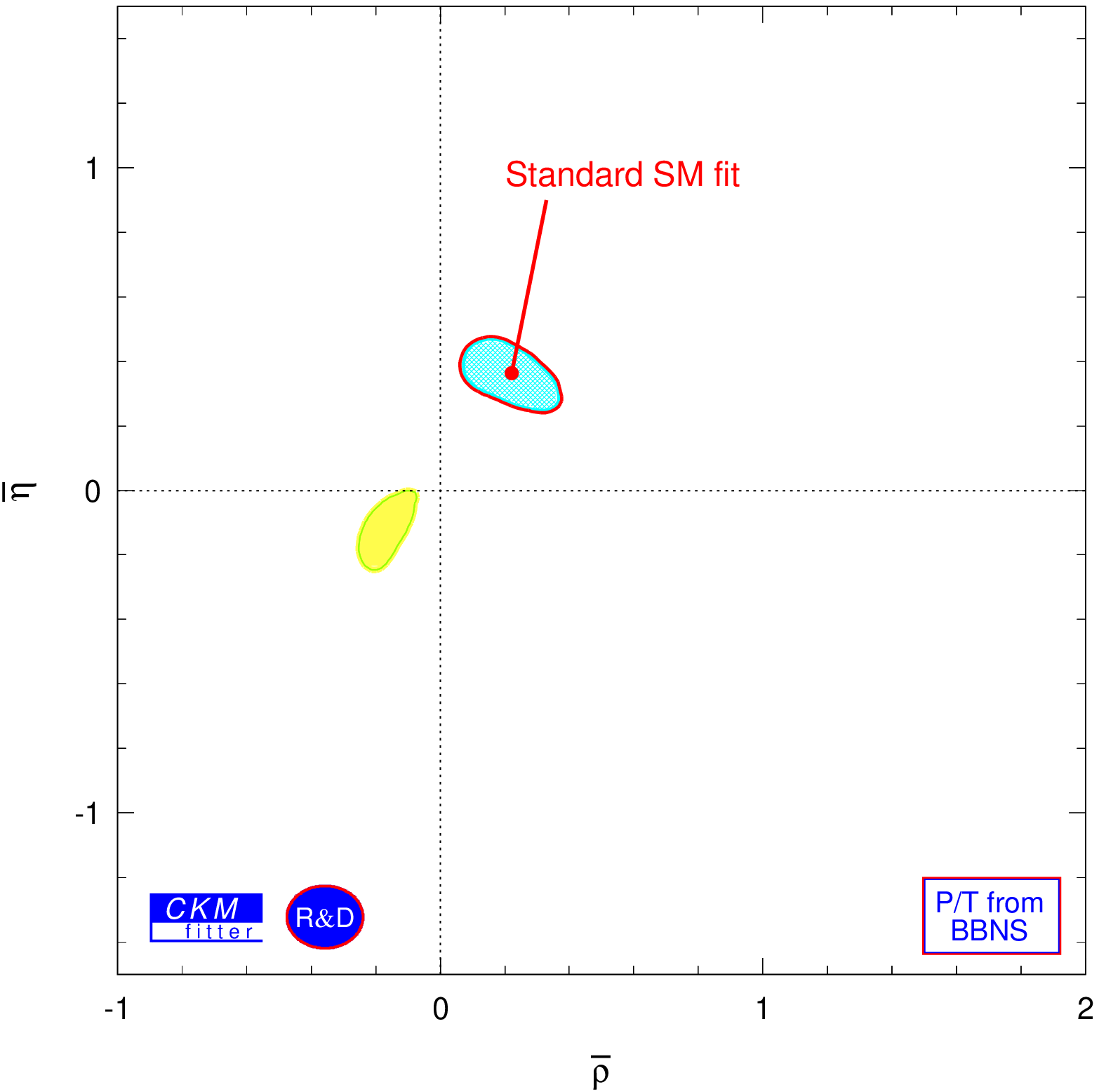,height=2.9in}
\psfig{figure=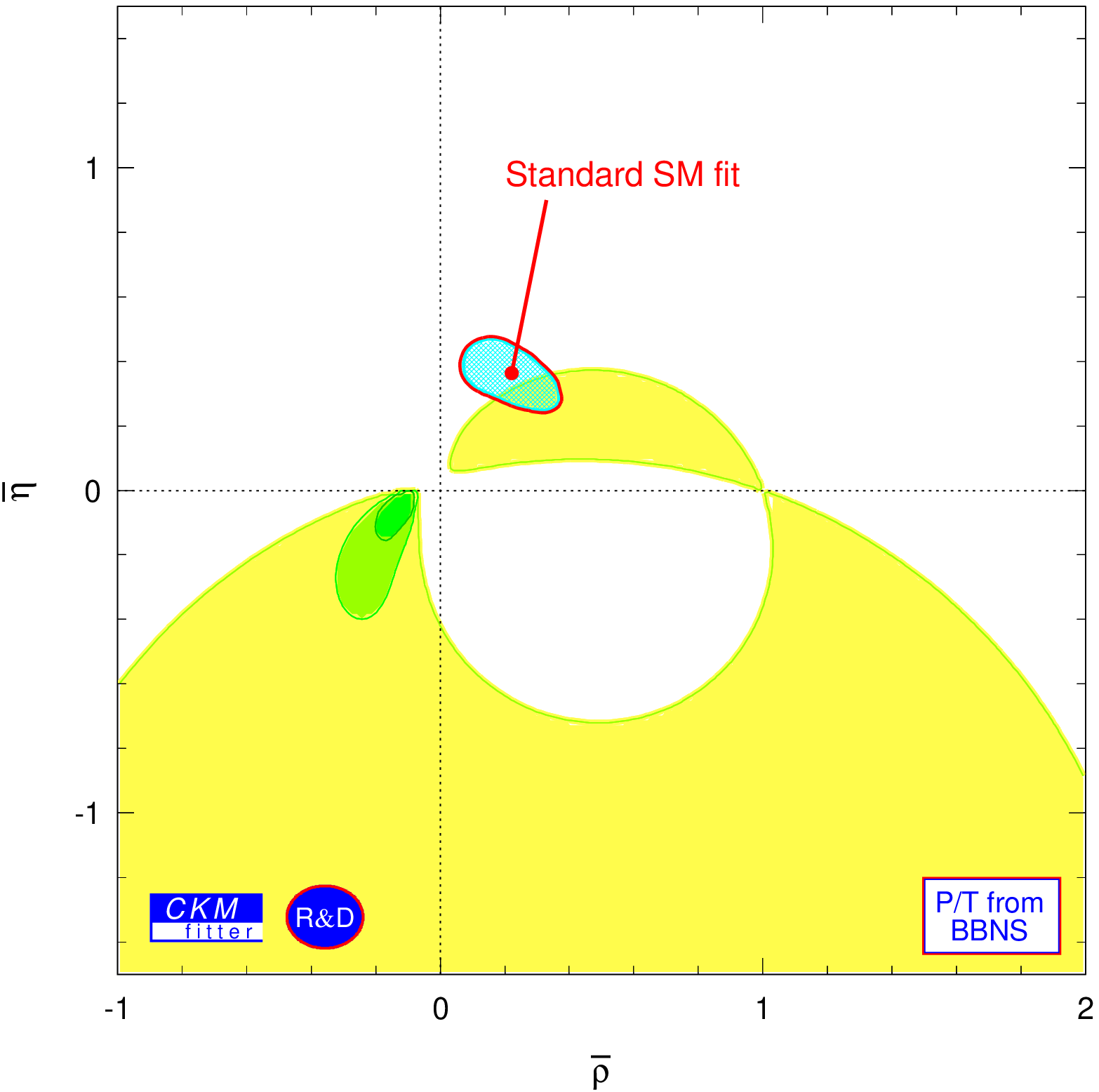,height=2.9in}
\caption{\underline{Left Plot}: CL in the $\rhobar-\etabar$ plane obtained from a fit using $\Cpipi$ and
$\Spipi$ from Belle, where the relative strong phase $\delta$ and the ratio $|P/T|$ are predicted by the QCD FA approach.
The light, medium and dark shaded zones include points with more than $5 \%$,$32 \%$,$90 \%$ CL. 
Also shown is the $95 \%$ CL zone of the SM fit.
\underline{Right Plot}: zoom of the CL between $0$ and $10 \%$.
\label{fig:cs}}
\end{center}
\end{figure}

The right plot in Fig.~\ref{fig:acp} and both plots in Fig.~\ref{fig:cs} show CL in the $\rhobar-\etabar$ plane
using the $\Cpipi$ and $\Spipi$ measurements of \babar\ and Belle respectively. While the \babar\ measurement,
interpreted within the QCD FA approach, is in good agreement with the SM expectations, the Belle measurement
is in strong disagreement. 
\section{Beyond the Standard Model: the example of $A_{\rm SL}$}
Two classes of models beyond the SM have been investigated~\cite{ASL}
using the semi-leptonic asymmetry which measures $CP$ violation in $B_d$-mixing:
\begin{eqnarray}
\nonumber
A_{\rm SL} &=& \frac{\Gamma(\bar{B^0}(t)\rar l^+ X) - \Gamma(B^0(t)\rar l^- X)}
                  {\Gamma(\bar{B^0}(t)\rar l^+ X) + \Gamma(B^0(t)\rar l^- X)} 
	    = {\rm Im}\frac{\Gamma_{12}}{M_{12}}.
\end{eqnarray}
 The present world-average~\cite{aSLmeas} reads $A_{\rm SL} = (0.2 \pm 1.4) \times 10^{-2}$. 
This parameter is predicted to be small (of the order
of $10^{-3}$) within the SM, but can be easily enhanced by new physics contributions.

In a class of models beyond the SM in which the CKM matrix is still a $3\times 3$ unitary matrix
and in which tree processes are dominated by the SM contributions (thus $\Gamma_{12} = \Gamma_{12}^{\rm SM}$,
and only $M_{12}$ is affected by new physics), $M_{12}$ can be parameterized by
\begin{equation}
M_{12} = r_d^2 e^{i2\theta_d} M_{12}^{\rm SM},
\end{equation}
where $r_d$ and $\theta_d$ are a modulus and a phase respectively.
Constraints from $\sin 2\beta$, $\Delta M_d$ and $A_{\rm SL}$ are thus also modified. 

\begin{figure}[h]
\begin{center}
\psfig{figure=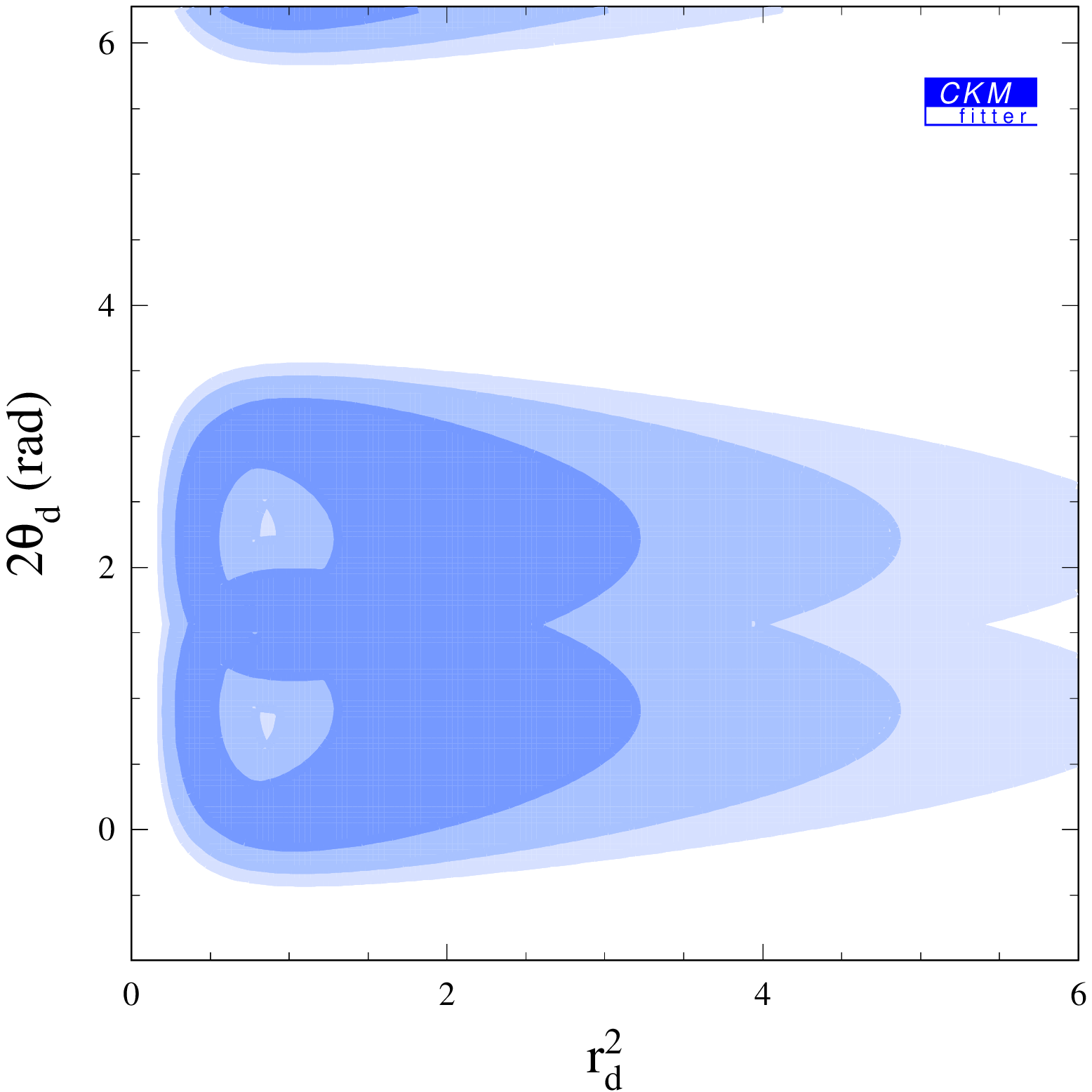,height=2.05in}
\psfig{figure=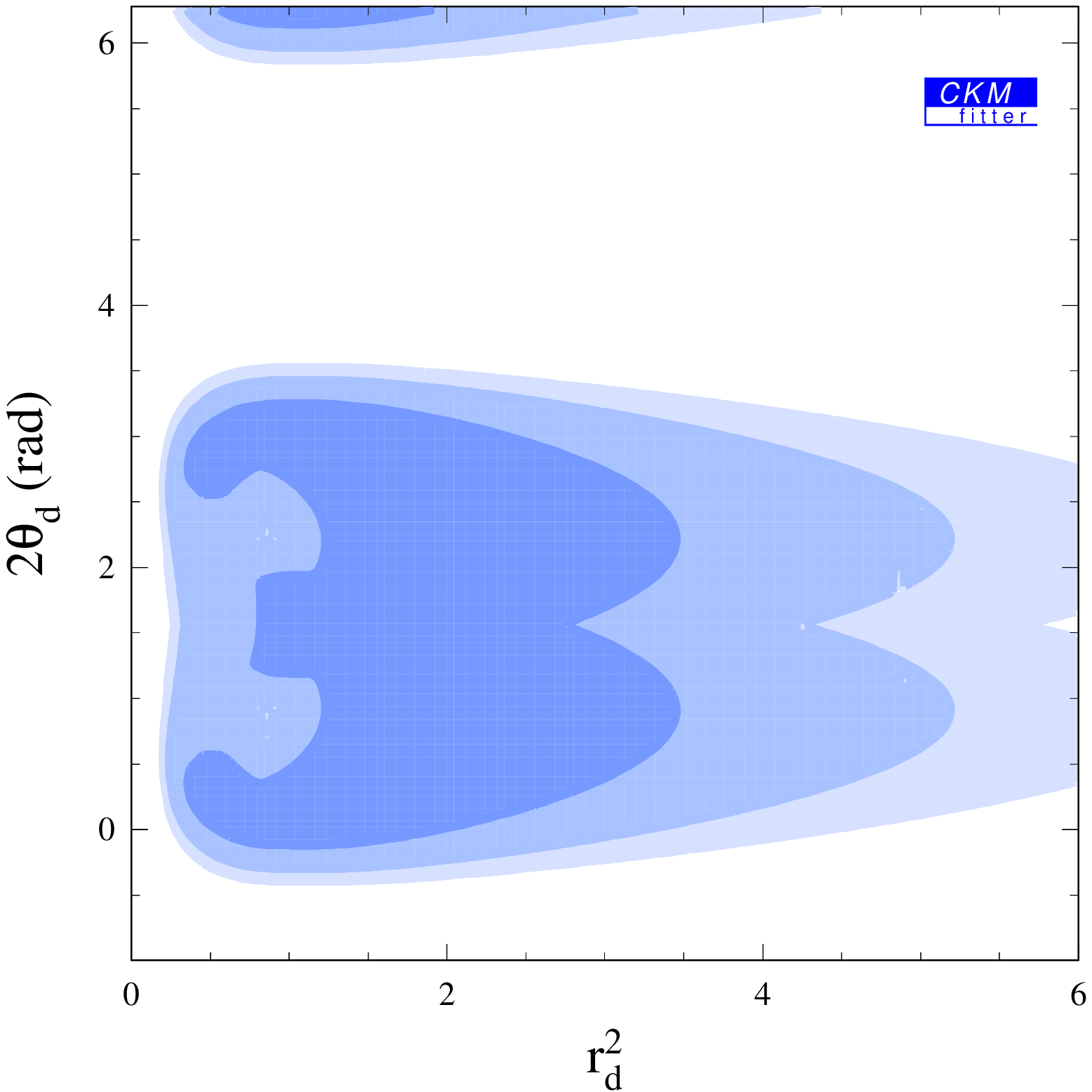,height=2.05in}
\psfig{figure=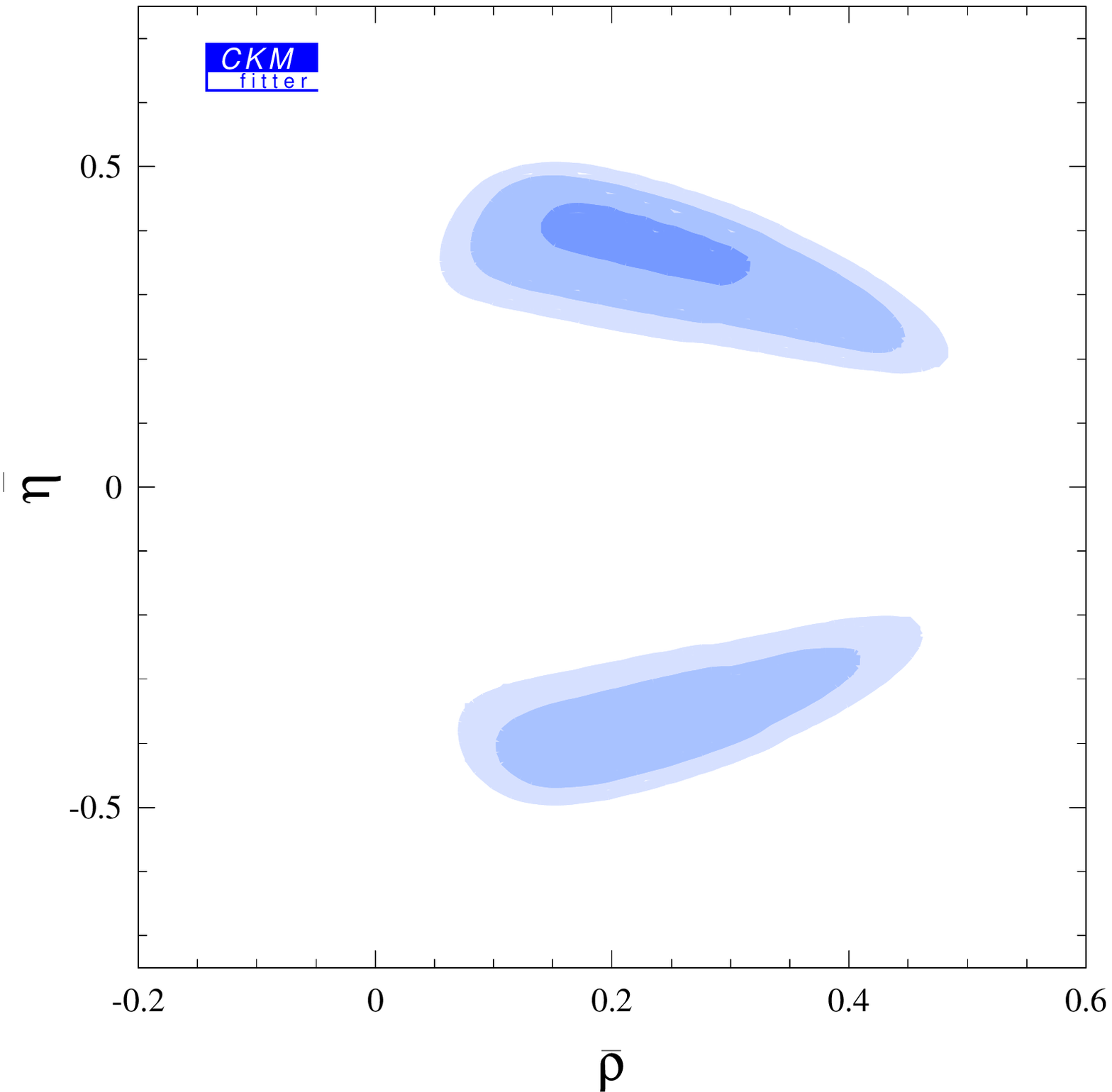,height=2.05in}
\caption{\underline{Left Plot}: CL in the $r^2-2\theta_d$ plane using the present standard
constraints of the $B_d$ system. The light, medium and dark shaded zones corresponds to $CL > 10$,
$32$ and $90 \%$ respectively.
\underline{Middle Plot}: CL in the $r^2-2\theta_d$ plane using the standard
constraints of the $B_d$ system as well as the $A_{\rm SL}$ measurement. \underline{Right Plot}: CL in 
the $\rhobar-\etabar$ plane in the framework of the MFV models. Negative values of $\etabar$ are
allowed.
\label{fig:asl}}
\end{center}
\end{figure}

The two left plots of Fig.~\ref{fig:asl} show CL in the ($r_d^2 - 2\theta_d$) plane using all present constraints
of the $B_d$ meson system, including (middle plot) or not (left plot) the constraint from
$A_{\rm SL}$: one sees that the $A_{\rm SL}$ constraint disfavours the small $r_d^2$ and large $\sin2\theta_d$
values.

A particular subset of the previous class of models have no new phase in flavour changing 
processes: these are the Minimal Favour Violating (MFV) models, as the minimal
MSSM, the two-Higgs doublets model, etc. In these models, one has:
\begin{equation}
r_d^2 = \frac{|F_{\rm tt}|}{S_0(x_t)}~~,~~2\theta_d = 0(\pi),
\end{equation}
where $S_0(x_t)$ is the Inami-Lim function~\cite{Inami} depending on the top quark mass.
$F_{\rm tt}$ enters $\Delta M_d$, $\Delta M_s$, $\epsilon_K$,
$\sin 2\beta$ and $A_{\rm SL}$. The right plot of Fig.~\ref{fig:asl} shows CL in the ($\rhobar-\etabar$)
plane within the MFV framework, where negative values of $\etabar$ are no longer excluded~\footnote{This solution is highly
dependent on the $f_{Bd}$ value entering the $\Delta M_d$ prediction: it disappears for $f_{Bd}$
larger than $210~MeV$~\protect\cite{ASL}.}.
As can be seen through the example of $A_{\rm SL}$, a global CKM analysis is not only the place to test
the Standard Model CKM paradigm, but also to investigate New Physics models, on which one can start to put 
meaningful constraints. When measurements reach a better precision, one will be able to rule out
models, while favouring others.
\section{Conclusion}
The successful operation of the $B$ factories at SLAC and KEK
has provided a flood of new results that can be used to obtain
information on the $CP$-violating phase of the CKM matrix. 
The extraordinary agreement of the world average of $\stb$ with the 
indirect determination represents a new triumph for the Standard Model. 
It establishes the KM mechanism as the dominant 
source of $CP$ violation in flavour changing processes at the electroweak scale.
While the inclusion of $\stb$ into the global CKM fit is a theoretically straightforward task, 
exploiting rare $B$ decays requires some further theoretical understanding in order 
to lead to meaningful constraints at present. 
{\small
\section*{Acknowledgements}
Besides the one who was luckily present at the Moriond conference,
four more persons are behind the \ckmfitter~group appellation and thus largely 
contributed to this presentation: A.~H\"ocker, H.~Lacker, F.~Le~Diberder, and J.~Ocariz. 
I am indebted to all of them for such an enjoyable working atmosphere. 
I would like to thank as well the conference organizing committee for the invitation,
the financial support, and the remarkable organization.
}


\begin{thebibliography}{99}
\bibitem{ckmFits}    	Several groups have pioneered global CKM fits in the past. For the attempt of
			an exhaustive list, see Ref.~[8]
\end{thebibliography}
\end{document}
